\documentclass[prl,twocolumn,floatfix,showpacs,superscriptaddress]{revtex4-1} %

\usepackage{mathptmx} 

\usepackage{amsmath}
\usepackage{graphicx,epstopdf}
\usepackage[utf8]{inputenc}

\newcommand{\abs}[1]{\ensuremath{\left| #1 \right|}}

\newcommand{\vq}{\mathbf{q}}

\begin{document}

\title{Paradigm shift for quantum paraelectric: softening of longitudinal modes}

\author{Yaron Kedem}
\affiliation{Department of Physics, Stockholm University, AlbaNova University Center, 106 91 Stockholm, Sweden}

\begin{abstract}
We revisit the quantum phase transition from a paraelectric state to a ferroelectric one and in particular the widespread distinction between a longitudinal modes to transverse one. In contrast to transitions at finite temperature, for a quantum phase transition breaking a discrete symmetry the splitting between the modes is irrelevant. We show that an anisotropy in the context of a quantum phase transition leads to a different behavior, compared to a classical transition, and suggest an experiment to observe it. The result is essential for explaining the pairing mechanism in strontium titanate.

\end{abstract}

\maketitle

Often, when physicists use the term ``quantum superposition” they refer to the state of photons, electrons or magnetic spins and not to the position of heavy ions in a crystal. Quantum paraelectrics are the exception to this rule \cite{quantPara,quantFluct}. These materials are on the verge of Ferroeletricity (FE) but even if one can cool them, to remove thermal fluctuations, still remain in the symmetric paraelectric phase. A prominent example is strontium titanate (STO) that can become FE by strain \cite{quantPara,stressIn}, Ca doping \cite{Ca} or isotope replacement \cite{Itoh1999}. There is much theoretical \cite{softTheo} and experimental \cite{expRaman} work investigating this transition. Recently, Rowely et al. \cite{rowley} showed the system goes through a quantum phase transition (QPT) manifesting quantum criticality in the context of FE. There is also a vast interest in how this phenomenon interplay with other phases such as superconductivity \cite{JonaPrl,FEinDome} and magnetism \cite{multi1,multi2}.  

The theory of FE poses many thorny physical challenges \cite{FE}. A ferroelctric material does not create a macroscopic field that can be measured directly like its magnetic analogue, making the definition and modeling of the phenomenon quite subtle. It is sensible to assume a magnetic dipole with a definite orientation is located in each unit cell, even in the absence of an external field. In contrast, the direction of an electric dipole depends on the arbitrary definition of the unit cell. Instead, one typically employs a description in terms of a Berry phase, focusing on the change in polarization, a quantity that can be measured. The different definitions affect the type of microscopic models one can imagine. Rather than of considering a collection of dipoles it seems more reasonable to focus on the motion of ions as a response to an external field. This have led to the theory of soft mode by Cochran \cite{cochran}, characterizing FE as a frozen optical phonon. At the phase transition the frequency of that phonon vanishes so even an infinitesimal electric field can polarize the system. The idea fits nicely with the relation Lyddane, Sachs and Teller have derived much earlier \cite{lst} 
\begin{align} \label{lst}
{\omega_{LO}^2 \over \omega_{TO}^2} ={\epsilon_{st} \over \epsilon_\infty } ,
\end{align}
where $\omega_{TO}$ and $\omega_{LO}$ are the frequencies of the transverse optical (TO) phonon and the longitudinal optical (LO) phonon, respectively and $\epsilon_{st(\infty)}$ is the static (high frequency) permittivity. This relation connects the divergence of susceptibility $\epsilon_{st} \rightarrow \infty$ to the vanishing frequency of the soft phonon 
\begin{align}  \label{softPhonon}
\omega_{TO} \propto \abs{T-T_c}^\beta,
\end{align}
 where $T$ is the temperature having a critical value $T_C$ for which the system turns FE. $\beta$ is a critical exponent denoting the universality class and often assumed to take the mean field value $\beta=1/2$ in the context of FE. 

A central property of a QPT is that the spectrum of the system becomes gapless at the critical point. So the energy of the lowest excitation vanishes as \cite{sachdevBook}
\begin{align} \label{gapVanish}
\Delta \propto \abs{g-g_c}^\nu,
\end{align}
where $g$ is a parameter with $g=g_c$ at the phase transition and $\nu$ is the critical exponent. It is then seems reasonable to identify $\omega_{TO} \equiv \Delta$ and replace $g \rightarrow T$ so one can apply the soft mode theory even when the transition is at zero temperature, possibly with some modification of the critical exponent. Indeed this was done extensively \cite{millis,softTheo}. Here, we argue that this identification and the subsequent application of the theory are not valid in general. While $\omega_{TO}$ and $\Delta$ are related, there are important properties that differ the mechanism of a QPT from the transition to FE at finite temperature. One such difference is that while the usual soft mode theory asserts that only $\omega_{TO}$ vanishes at the transition and that $\omega_{LO}$ is largely independent of temperature, at certain QPTs there is no clear distinction, implying $\omega_{LO}$ also softens.

There are three main issues we need to clarify in order to substantiate the argument: (i) How and when is the LO-TO splitting, which is usually considered universal, absent? (ii) What is the essential difference between QPTs and those at finite temperature, which share much of the phenomenology and theoretical treatment, prompting a different behavior for the longitudinal mode? (iii) Show that current experimental evidence for softening $\omega_{TO}$ do not rule out the softening of $\omega_{LO}$ and how the latter can be observed experimentally. Below we discuss each of these issues and then comment on the implications to superconductivity. 

\emph{LO-TO splitting:}
Before going into the mathematical treatment of the LO-TO splitting, we should specify what do we mean with ``transverse” and ``longitudinal”. At any finite crystal momentum $\vec{q}$, it is straight forward to define the motion of the ions in a direction parallel to $\vec{q}$ as longitudinal and the perpendicular as transverse. However we are interested in the limit $\abs{q}\rightarrow 0$. The mathematical difficulty which we will encounter is related to some subtlety of the definition at this limit. Instead of choosing an arbitrary direction for $\vec{q}$ and separating the ions displacement to two components, one can consider an arbitrary direction for the motion of the ions and separate the components of the momentum to the parallel and perpendicular directions. In an anisotropic material the latter choice might be more suitable. Note that much of the theoretical treatment in the literature consider the response of the system to an external field. In that case the longitudinal and transverse directions are set a priori and this issue is irrelevant. Conversely, we are interested in the autonomous dynamics of the system.

The LO-TO splitting is usually considered to come from long range dipole-dipole interaction. In general, such interactions do not allow analytical solutions nor reliable numerical calculations, so one have to use physical arguments and make qualitative statements. The interaction between two electric dipoles depends on the displacement between them, on the angle between their orientations and on the angles between the displacement and each orientation. Fourier transforming from displacement to crystal momentum, one obtains a term
\begin{align} \label{inter}
V_{a b}(\vec{q}) \propto {q_a q_b \over q^2},
\end{align}
where $a$ and $b$ denote the orientations of the two dipoles. The limit $\abs{q}\rightarrow 0$ can yield two different results for $V_{a b}(0)$, depending on whether $\vec{q}$ has $a$ and $b$ components. For a certain $\vec{q}$, a parallel motion entail an additional energy term, compared to perpendicular motion, even as we take $\abs{q}\rightarrow 0$. The extra term induces the LO-TO splitting, which is present in many cases. 

Consider now a strongly anisotropic system such that some ions can move only along one axis, say $z$. The system is 3D so Eq. (\ref{inter}) can still describe the interaction between the dipoles created by the motion of the ions. There can be other short range interactions and kinetic energy as well. Studying the frequency of the modes $\omega_\vq $ at small $\abs{q}$, we are likely to find much stronger dispersion in the $z$ direction than along $x$ and $y$. However, any finite contribution $\omega_0$ left at the limit $\abs{q}\rightarrow 0$ cannot be assigned to the LO-TO splitting, as there is only one possible polarization and at $\abs{q} = 0$ it cannot be labeled to as ``transverse” or ``longitudinal”. Note that assuming $\omega_\vq $ is non analytic at $\abs{q} = 0$ so $\omega_0$ might not be unique, is not enough to recover the splitting. One can consider a finite transverse component $q_x$ and take the limit of vanishing longitudinal component $q_z \rightarrow 0$, or vice versa, interchanging LO and TO. A non unique $\omega_0$ implies extending the non analytical behavior to the $q_z = 0$ plane.   

When the system goes through a QPT, $\omega_0$ has to vanish (assuming the transition is at zone center and not in some finite $\vq$). Since the transverse dispersion is much weaker there will be a larger range for $q_x$ and $q_y$ where  $\omega_\vq $ is small, compared to the range for $q_z$. At any finite  $\vq $ we can assign the LO-TO splitting to the dispersion in different direction but not to the vanishing $\omega_0$ at the QPT.    

We can now relax slightly the strong anisotropy assumption and allow the ions to move along the other axes. We treat the motion in each direction as a separate mode and denote its frequency as $\omega^\alpha_\vq $ with $\alpha = x,y,z$. The modes of different direction might interact and modify the exact expression for $\omega^\alpha_\vq $ but as long as the system is not isotropic we cannot treat them as a common degree of freedom. Most importantly, since FE does not occur at an arbitrary direction but along one of the preset axes, the physical picture above, focusing on motion in one direction, is valid at the transition.
        
\emph{QPT vs. finite temperature:}     
A prototypical model for a QPT is to consider a quantum rotor at each lattice point \cite{sachdevBook}. The Hamiltonian is give by
\begin{align} \label{rotor}
H_R=\frac{J g}{2}\sum_i L_i^2 - J \sum_{\langle ij\rangle}\hat{n}_i\cdot\hat{n}_j,
\end{align}
where $J$ is the energy scale, $\hat{n}$ is an operator representing a unit vector in $N$ dimensions and $L$ is an operator for the kinetic energy (angular momentum).  For $N>1$ this Hamiltonian has a continuous $O(N)$ symmetry, while for $N=1$ one can replace $\hat{n}$ and $L^2$ with Pauli matrices to obtain the quantum Ising model with a discrete $Z_2$ symmetry. In both cases, when $g<g_c$ the ground state breaks the symmetry and the system has a gap $\Delta \propto J (g_c-g)^\nu$. Above some critical temperature $T_c \propto \Delta$, thermal excitations restore the symmetry, but strictly speaking, the spectrum of a Hamiltonian does not change and in particular $\Delta$ is temperature independent. From this perspective, equations (\ref{softPhonon}) and (\ref{gapVanish}) describe utterly different mechanisms.

It seems strange that two effects sharing much of their phenomenology would be completely unrelated. Moreover, while the microscopic spectrum is in principle temperature independent, the macroscopic dynamics are certainly affected by the temperature, especially close to a phase transition. Indeed, we can use the Hamiltonian in  Eq. (\ref{rotor})  to describe a phase transition at finite temperature, or equivalently write a field theory where one can do a Wick rotation to finite temperature. As we mentioned above for $T_c>0$, the symmetry is broken in the ground state and there is a gap $\Delta>0$. However, for $N>1$ the broken $O(N)$ symmetry implies there are $N-1$ gapless transverse modes, i.e. Goldstone bosons. Their existence does not depend on temperature but at certain temperature they will dominate the response of the system. 
 
\begin{figure*}
\includegraphics[trim=3cm 4cm 2cm 5.3cm, clip=true,width=0.99\textwidth]{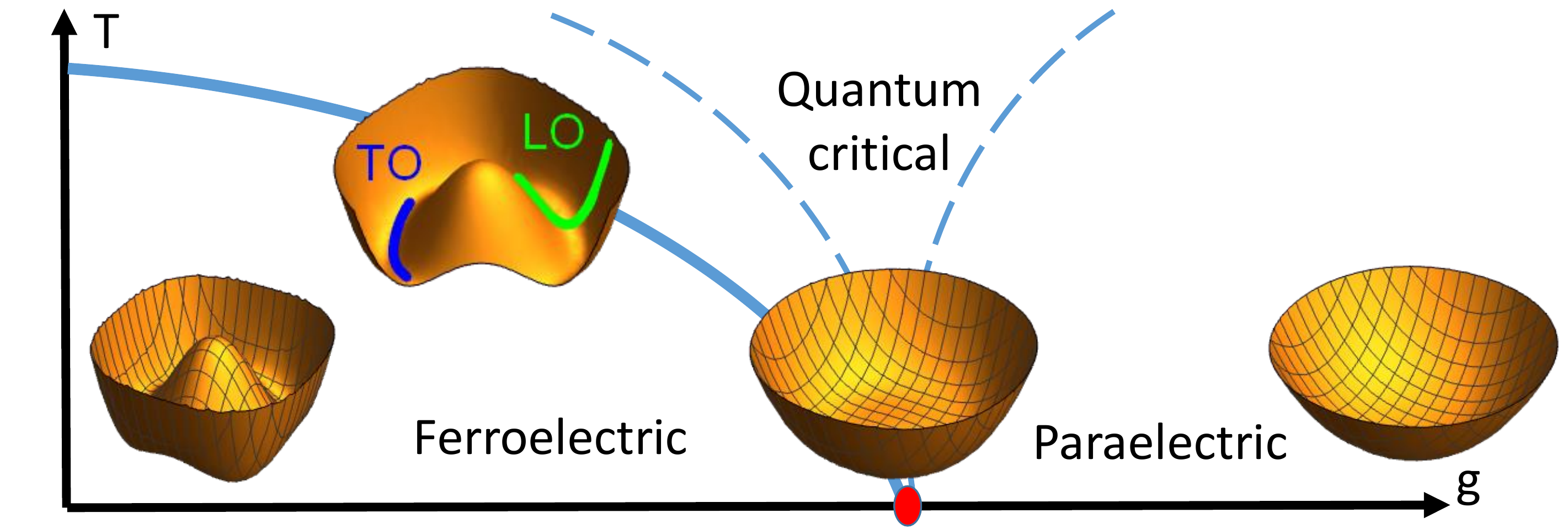}
\caption{A typical phase diagram of a ferroelectric quantum critical point (QCP), showing the potential of an effective field theory based on the Hamiltonian in Eq. (\ref{rotor}) and a perturbation similar to Eq. (\ref{latt}) with $N=2$. Tuning $g$ modifies the Hamiltonian and the resulting potential from having a symmetric ground state for large $g$ (on the right side) to one with a symmetry breaking ground state for small $g$ (on the left side) via a potential with a flat minimum (in the middle). The perturbation break explicitly the rotational symmetry so only a discrete one is broken spontaneously. When the transition is at finite temperature the state sinks into the ``Mexican hat''. The longitudinal modes retain a large gap while the transverse ones only have a gap due to the perturbation and appear soft. }
\label{phase}
\end{figure*}

Let us now try to consider how this mechanism would look in a realistic material. In contrast to a magnetic dipole, which plausibly have an $O(N)$ symmetry, FE is closely related to the lattice structure so only discrete symmetries are possible in the microscopic level. However, on the macroscopic level at finite temperature a cubic $O_h$ symmetry might be washed out to an effective spherical $O(3)$ symmetry and similarly a square $D_4$ to a circular $O(2)$. We can model this effect, for example in the case of $N=3$, by adding to the Hamiltonian in Eq. (\ref{rotor})  a term
\begin{align} \label{latt}
H_c= \tilde{J} \sum_i \abs{\vec{P_1} \cdot \hat{n}_i} +  \abs{\vec{P_2} \cdot \hat{n}_i}
\end{align}
where $ \tilde{J} \ll J$ is a much smaller energy scale and $\vec{P_{1,2}}$ are two vectors in 3 dimensions representing the effect of the crystal structure. The new Hamiltonian $H_R+H_c$ has a $Z_2$ symmetry for $\hat{n} \rightarrow - \hat{n}$ and if $\vec{P_1} \cdot \vec{P_2} =0$ another two $Z_2$ symmetries for reflections (other forms of $H_c$ can yield a variety of symmetry groups), but there is no continuous symmetry. However, when $g\ll g_c$, the effect of $H_c$ might be negligible. The transverse Goldstone modes of $H_R$ will obtain a gap (``mass") $\propto \tilde{J}$ that is much smaller than the gap $\propto J$ of the longitudinal mode and might be hardly observable in experiment. Thus, we expect the approximated Goldstone modes to start dominate when the temperature is lowered to the vicinity of the gap of $H_R$, showing a soft response together with a breaking of the continuous symmetry. At even lower temperatures the small mass will become significant and the system hardens again while breaking a discrete symmetry.  

Closer to the critical point, where $H_R$ becomes completely gapless, the effect of $H_c$ cannot be neglected anymore. One can still have a phase transition, which in general might be at a different value $g = \tilde{g_c}$, but with no Goldstone mode as the broken symmetry is now discrete. The idea is illustrated in Fig. \ref{phase}.

The model in Eq. (\ref{rotor}) is sometime referred to as describing a transition of the type ``order-disorder'', while FE transitions are usually considered to be ``displacive'' \cite{disp}. From macroscopic characteristics such as $\epsilon_{st}$, it is difficult to distinguish between the two types of transition. The microscopic physics should differentiate between them but for a quantum model this distinction does not apply straightforwardly. While the operators in Eq. (\ref{rotor}) have discrete eigenvalues, their expectation value is continuous. Even if one starts by writing a model in terms of continuous variables such as position and momentum, a proper treatment involves finding a discrete spectrum, along with the corresponding operators (these would be the ladder operators for an Harmonic potential but can be defined more generally). 

One might be skeptical to the use of the model in Eq. (\ref{rotor}), as ferroeceltric dipoles, in contrast to magnetic ones, are not well defined objects on the atomic level. The operators of our model refer to certain states of the spatial configuration of ions in a unit cell. The separation to unit cells is indeed arbitrary and one can define other operators based on a different partition to unit cells. Some quantities in the redefined model might change but the critical point will not. Regardless of the partition there is entanglement with neighboring unit cells due to the interaction and the state of the system is not an eigenstate of a single unit cell operator. Indeed even if one knows the exact state of the system there is no definite way to say whether a electric dipole in a specific unit cell is pointing in a particular direction.

\emph{Experimental observations:} 
The discussion above is general but in an experimental context, STO is the prominent or sole example of a quantum paraelectric. As shown above one should differentiate transitions at finite temperature and a QCP where tuning some parameter in the Hamiltonian make the spectrum gapless. Moreover, not all tuning parameters can bring the system to the QCP. Replacing some strontium atoms with calcium \cite{Ca} creates a local polarization at the calcium site and the transition to FE is likely to be percolative. Uniaxial strain creates an additional anisotropy imposing an external distinction between TO and LO. A QPT is likely to occur when one change the isotopic mass \cite{Itoh1999}, which does not directly couple to electric polarization and is typically done in a rather homogeneous fashion. Regardless of how one can drive the transition, pristine STO is extremely close to the QCP so the LO are likely to be already soft.   
  
Some experimental and numerical results present the values of $\omega_{TO}$ and $\omega_{LO}$ \cite{mech}. Typically there are several values corresponding to different modes. It was also shown that the value of the lowest $\omega_{TO}$  decreases as the temperature is lowered, indicating the softening of the modes. A vast majority of the experiments are using light, for example Raman spectroscopy \cite{expRaman} or reflectivity measurements. However, the electric field in light is transverse and does not couple directly to LO. For a given model one can infer the values of $\omega_{TO}$ and $\omega_{LO}$ but such experimental results do not invalidate a different model yielding a different value for $\omega_{LO}$. An alternative method is neutron scattering. There, both the TO and LO should impact the result, but it is not trivial to distinguish their contributions. Some neutron scattering revealed a violation of Eq. (\ref{lst}) \cite{LSTviolate1,LSTviolate2}.

\begin{figure}
\includegraphics[trim=3cm 7.8cm 4cm 3cm, clip=true,width=0.48\textwidth]{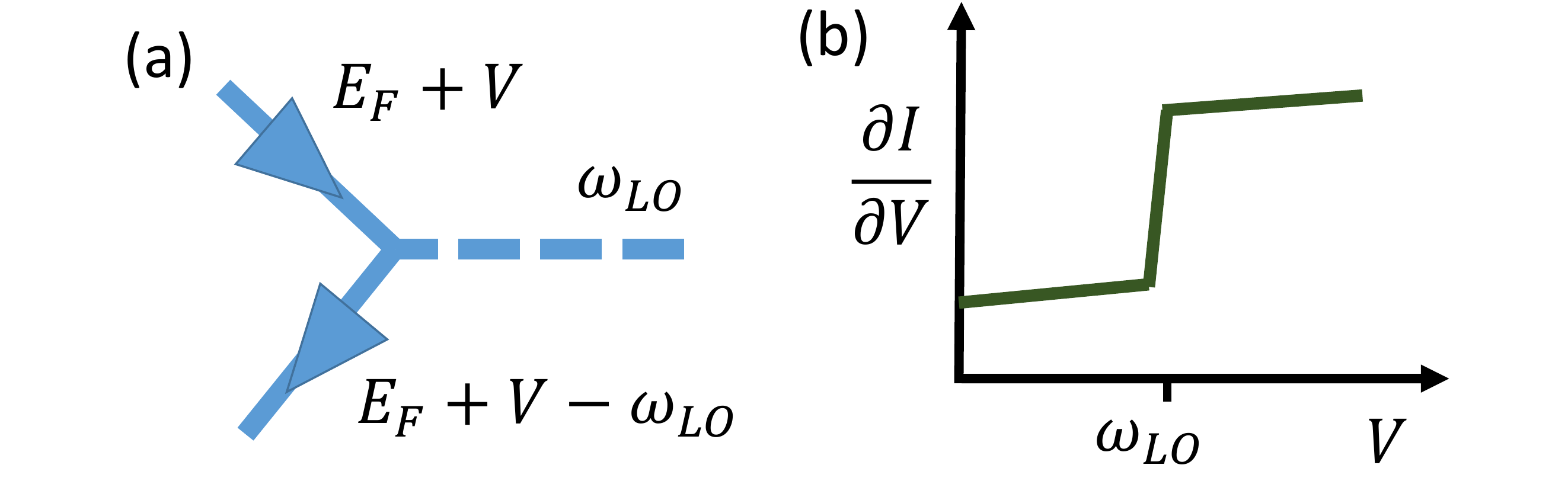}
\caption{Experimental observation of LO softening: (a) An electron with a bias $V$ scatters a LO and loses energy $\omega_{LO}$ before entering the band. (b) Using a scanning tunneling microscope (STM) one can sweep $V$ and measure the current $I$. The derivative $\partial I / \partial V$ is proportional to the density of states in the band at the energy of the electron. Tuning the material closer to the QCP will shift the step closer to zero bias.}
\label{stm}
\end{figure}

There is another possible probe besides photons and neutrons: electrons. In inelastic electron tunneling spectroscopy, scattering with a phonon can shift the energy level of the electron. The result is a kink in the I-V curve, or a step in its first derivative, at the frequency of the phonon. Since electrons couple to LO rather than to TO such a process would indicate the value of $\omega_{LO}$ as shown in Fig. \ref{stm}. As one tune the material closer to the QPT, our model predicts the position of the step will move closer to zero bias. 

\emph{Implications to superconductivity:} Until now we have discussed only the physics of the FE transition on its own. It can be important to understand the interplay of this transition with other phases. In particular, we suggest the LO modes play a pivotal role in the pairing mechanism of low doping superconductivity. For a low density of carriers, one needs a long range interaction to couple electrons. In other words, when the Fermi surface is small the interaction has to be strong at small momenta. Structural modes related to FE mediate an effective interaction between electrons \cite{novel}
\begin{align} \label{veff}
V_{eff}(\vq) \propto  \sum_\alpha {g^\alpha_\vq \over  \omega^\alpha_\vq}
\end{align}
where the coupling $g^\alpha_\vq \propto q_\alpha \abs{\vq}^{-2}$ between these modes and the electrons is peaked around $q=0$ but vanishes for purely transverse momentum, i.e. when $\abs{\vq} \ne 0, q_\alpha = 0$. Above we argued that close to the QCP, $\omega^\alpha_\vq$ is vanishing at $q=0$ and has steep dispersion in the $\alpha$, i.e. longitudinal, direction. These properties yield $V_{eff}(\vq) $ that is sharply peaked at $q=0$ and give rise to forward scattering \cite{forward1,forward3}, making this type of interaction highly suitable for low doping superconductivity.
  
For STO, the previous understanding that LO do not soften at the FE transition together with the observation that TO do not couple to electrons, led many authors to reject or ignore the connection between the corresponding QCP and superconductivity \cite{ruhman,gor2016phonon}. The large and anomalous isotope effect, which was predicted in \cite{JonaPrl,prbIso} and later observed in \cite{stucky}, is another evidence for the softening of LO (An alternative explanation can be a coupling to TO modes \cite{TO}). Even when the connection of superconductivity to the critical behavior is accepted, insisting that the LO remain hard leads to a number of problems when one wishes to explain the mechanism of superconductivity in STO \cite{mech,SCFEUltra,SCSTOrev}. Considering an interaction as in Eq. (\ref{veff}), with $\omega^\alpha_\vq$ vanishing at $q=0$ and rising sharply for finite $q_\alpha$, one obtain a forward scattering process that can lead to a Cooper instability even at vanishing doping level \cite{novel,instab}.  
 
We have critically analyzed the assumption that only TO modes become soft at a FE transition and showed that it is not valid for a QPT. A close inspection of the distinction between TO and LO in an anisotropic system revealed their splitting cannot be associated with a gap at the QCP. We have clarified the difference between a transition at finite temperature and the situation close to the QCP, in particular the role of anisotropy in determining the type of symmetry that is broken. The softening of LO does not conflict with any experimental results (that we are aware of) and we have proposed an experimental method to observe it. In addition to illuminating the nature of quantum paraelectrics, the result have important implications regarding the mechanism of superconductivity in STO. The new understanding is likely to shed light on the interplay of FE with other phases as well.

\begin{acknowledgments}
 I gratefully acknowledge useful discussions with Nicola Spaldin. This work was supported by the Swedish Research Council under Contracts No. 335-2014-7424, 2016-04192$\_$3, 2018-06720$\_$3 
\end{acknowledgments}

\bibliography{library_FELOTO}

\end{document}